\documentclass[showpacs,twocolumn,pra]{revtex4}
\usepackage{graphicx}
\usepackage{amsfonts}
\usepackage{amsmath}
\usepackage{amssymb}
\usepackage{mathrsfs}

\begin{document}

\title{Tight Upper Bound Of The Maximum Speed Of Evolution Of A Quantum State}

\author{H.~F. Chau}\email{hfchau@hkusua.hku.hk}
\affiliation{Department of Physics and Center of Computational and Theoretical
 Physics, University of Hong Kong, Pokfulam Road, Hong Kong}
\date{\today}

\begin{abstract}
 I report a tight upper bound of the maximum speed of evolution from one
 quantum state $\rho$ to another $\rho'$ with fidelity $F(\rho,\rho')$ less
 than or equal to an arbitrary but fixed value under the action of a
 time-independent Hamiltonian.  Since the bound is directly proportional to the
 average absolute deviation from the median of the energy of the state
 ${\mathscr D}E$, one may interpret ${\mathscr D}E$ as a meaningful measure
 of the maximum information processing capability of a system.
\end{abstract}

\pacs{03.65.-w, 03.67.-a, 89.70.Eg}

\maketitle

\section{Motivations And Prior Arts \label{Sec:Motivation}}
 It is impossible to build an arbitrarily fast and powerful computer, quantum
 or classical, because several fundamental physical limits bound the maximum
 speed of logical operations and the size of memory space~\cite{phy_limit}.  In
 particular, Bhattacharyya~\cite{delta_E1}, Uhlmann~\cite{delta_E2} and
 Pfeifer~\cite{delta_E3} found that the time $\tau$ needed to evolve a (mixed)
 state $\rho$ to another state $\rho'$ under that action of a time-independent
 Hamiltonian $H$ is tightly lower-bounded by
\begin{equation}
 \tau \geq \tau_\text{\tiny TEUR} \equiv \frac{\hbar \cos^{-1} \left(
 \sqrt{\epsilon} \right)}{\Delta E} \equiv \frac{g_\text{\tiny TEUR}(\epsilon)
 \pi\hbar}{2\Delta E} ,
 \label{E:TEUR_bound}
\end{equation}
 where $\epsilon = F(\rho,\rho') \equiv \left[ \text{Tr} ( \sqrt{\sqrt{\rho}
 \rho' \sqrt{\rho}} ) \right]^2$ is the fidelity between the two states and
\begin{equation}
 \Delta E = \sqrt{\text{Tr} ( H^2 \rho  ) - E^2} \equiv \sqrt{\text{Tr} ( H^2
 \rho ) - \left[ \text{Tr} ( H \rho ) \right]^2}
 \label{E:Delta_E_def}
\end{equation}
 is the standard deviation of the energy of the system.  (Actually,
 Refs.~\cite{delta_E2} and~\cite{delta_E3} considered the more general
 situation of a time-dependent Hamiltonian whose results can be reduced to
 Eq.~(\ref{E:TEUR_bound}) in the time-independent case.)  Because of the form
 of Eq.~(\ref{E:TEUR_bound}), it is sometimes called the time-energy
 uncertainty relation (TEUR) bound.  Bounds of this type are interesting for
 they depend only on a modest description of the system.  Later on, Margolus
 and Levitin~\cite{PHYSCOMP96,max_speed} discovered another tight lower bound
 on the time required to evolve a (pure) state to another state in its
 orthogonal subspace under the action of a time-independent Hamiltonian $H$.
 Their bound is inversely proportional to the average energy of the system
 above the ground state, $E - E_0$.  Giovannetti \emph{et al.}~\cite{q_limit}
 extended the Margolus-Levitin theorem by showing that the time $\tau$ required
 to evolve between two (mixed) states with fidelity less than or equal to a
 fixed $\epsilon\in [0,1]$ under the action of a time-independent Hamiltonian
 is tightly lower-bounded by
\begin{equation}
 \tau \geq \tau_\text{\tiny ML} \equiv \frac{g_\text{\tiny ML} (\epsilon) \pi
 \hbar }{2 (E - E_0)}
 \label{E:ML_bound}
\end{equation}
 for some smooth function $g_\text{\tiny ML}$.  Although no closed form
 expression is known for $g_\text{\tiny ML}$, it can be approximated to within
 a few percent of error by~\cite{q_limit}
\begin{equation}
 g_\text{\tiny ML} (\epsilon) \approx \left[ g_\text{\tiny TEUR}(\epsilon)
 \right]^2 .
 \label{E:g2_def}
\end{equation}
 More importantly, Giovannetti \emph{et al.} found examples in which the ML
 bound in Eq.~(\ref{E:ML_bound}) is better than the TEUR bound~\cite{q_limit}.
 Note that the smaller the $\tau$, the faster the system can be used for
 quantum information processing.  In this respect, the tight bounds in
 Eqs.~(\ref{E:TEUR_bound}) and~(\ref{E:ML_bound}) show that $\Delta E$ and $E -
 E_0$ are reasonable measures of the maximum possible quantum information
 processing rate of a system~\cite{phy_limit,PHYSCOMP96,max_speed,q_limit}.

 Here I report another tight lower bound on the time needed to evolve from one
 (mixed) state to another under the action of a time-independent Hamiltonian
 such that the fidelity is less than or equal to a fixed value $\epsilon \in
 [0,1]$.  Recall from the discussions of Margolus and Levitin in
 Refs.~\cite{PHYSCOMP96,max_speed} that the faster the time $\tau$ for a
 quantum system to evolve between two orthogonal states, the more powerful the
 system can process quantum information.  In this respect, a lower bound of the
 time $\tau$ poses a so-called quantum speed limit on the maximum quantum
 information processing rate of the system.  This notion of quantum speed
 limit was then generalized by Giovannetti \emph{et al.} to the study of
 evolution between two non-orthogonal states~\cite{q_limit}.  Since the new
 tight evolution time bound reported here is inversely proportional to the
 so-called average absolute deviation from the median (AADM) of the energy of
 the state ${\mathscr D}E$ of the system, I conclude that ${\mathscr D}E$ is
 also a reasonable measure of the maximum possible quantum information
 processing rate of a system.  Finally, I compare this bound with the TEUR
 bound~\cite{delta_E1,delta_E2,delta_E3} and the ML
 bound~\cite{PHYSCOMP96,max_speed,q_limit}.

\section{The New Evolution Time Bound \label{Sec:Bound}}
\subsection{An Auxiliary Inequality \label{Subsec:Lemma}}
\begin{figure}[t]
 \centering
 \includegraphics*[scale=0.8]{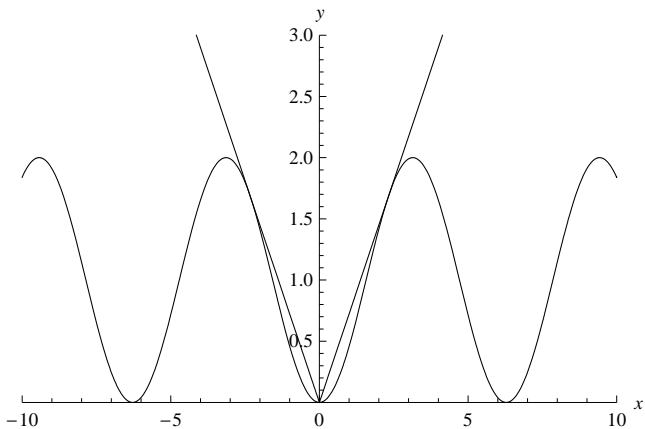}
 \caption{The curve $y = 1-\cos x$ and the broken line $y = A |x|$ defined in
  the text.
  \label{F:geometry}
 }
\end{figure}

 I begin by considering an inequality with a simple geometric meaning.  The
 first quadrant of Fig.~\ref{F:geometry} depicts the (unique) line with the
 greatest slope that passes through the origin and meets the curve $y = 1 -
 \cos x$ at two distinct points (namely, $x = 0$ and $x = x_m$).  Clearly, this
 line is the tangent to the curve at $x = x_m$; and its slope $A$ is given by
\begin{equation}
 A = \max \left\{ \frac{1 - \cos x}{x} : x > 0 \right\} .
 \label{E:A_def}
\end{equation}
 Numerically, I find that
\begin{subequations}
\begin{equation}
 A \approx 0.724611
 \label{E:A_numerical}
\end{equation}
and
\begin{equation}
 x_m \approx 2.33112 .
 \label{E:x_m_numerical}
\end{equation}
\end{subequations}

 By considering the mirror image of this line with respected to the $y$-axis,
 it is obvious that
\begin{equation}
 \cos x \geq 1 - A | x | \label{E:basic_inequality} 
\end{equation}
 for all $x \in {\mathbb R}$.  (The geometric meaning of this inequality is
 apparent from Fig.~\ref{F:geometry}.)

\subsection{The Pure State Case \label{Subsec:Pure_State}}
 Now, I may use Margolus and Levitin's argument in
 Refs.~\cite{PHYSCOMP96,max_speed} to obtain the required bound for the case of
 pure states.  Suppose $|\Phi (0)\rangle = \sum_j \alpha_j |E_j\rangle$ where
 $|E_j\rangle$'s are the normalized energy eigenvectors of the
 time-independent Hamiltonian $H$ and $\sum_j |\alpha_j|^2 = 1$.  Then under
 the action of $H$,
\begin{equation}
 \langle \Phi (0) | \Phi (t) \rangle \equiv \langle \Phi (0) | e^{-i H t /
 \hbar} | \Phi (0) \rangle = \sum_j |\alpha_j|^2 e^{-i E_j t / \hbar} .
 \label{E:evolution}
\end{equation}
 In other words, at the time when the system evolves to a state whose fidelity
 is less than or equal to $\epsilon$ from $|\Phi (0)\rangle$, the real part of
 Eq.~(\ref{E:evolution}) obeys
\begin{equation}
 \left| \sum_j |\alpha_j|^2 \cos \left( \frac{-E_j t}{\hbar} \right) \right|
 \leq \sqrt{\epsilon} .
 \label{E:real_part}
\end{equation}
 Applying the inequality in Eq.~(\ref{E:basic_inequality}) to
 Eq.~(\ref{E:real_part}), I get
\begin{equation}
 1 - \frac{A t}{\hbar} \sum_j |\alpha_j|^2 | E_j | \leq \sqrt{\epsilon} .
 \label{E:time_1}
\end{equation}
 Therefore, the earliest time $\tau$ at which $|\Phi (0)\rangle$ evolves to a
 state whose fidelity is less than or equal to $\epsilon$ from $|\Phi (0)
 \rangle$ satisfies the inequality
\begin{equation}
 \tau \geq \frac{\left( 1 - \sqrt{\epsilon} \right) \hbar}{A \sum_j
 |\alpha_j|^2 |E_j|} .
 \label{E:time_bound}
\end{equation}

 By means of the fact that the reference energy level of a system
 has no physical meaning, I can further strengthen the bound in
 Eq.~(\ref{E:time_bound}) as follows:  Recall that the function $f(x) = \sum_j
 |\alpha_j|^2 |E_j - x|$ attains its minimum when $x$ equals $M$, the median of
 $E_j$'s with relative frequency of occurrence of $E_j$ equals $|\alpha_j|^2$.
 More precisely, consider the cumulative distribution function
\begin{equation}
 C(x) = \sum_{j: E_j \leq x} |\alpha_j|^2 .
 \label{E:CF}
\end{equation}
 Then
\begin{equation}
 M = \frac{1}{2} \left[ \lim_{y\rightarrow 0.5^-} C^{-1}(y) +
 \lim_{y\rightarrow 0.5^+} C^{-1}(y) \right] .
 \label{E:M_def}
\end{equation}
 (The above assertion can be proven by checking when $df/dx = 0$.)  In
 statistics, the quantity
\begin{eqnarray}
 {\mathscr D}E & \equiv & \sum_j |\alpha_j|^2 |E_j - M| \nonumber \\
 & = & \text{Tr} \left[ \sqrt{ (H - M)^\dagger (H - M)} | \Phi(0)
  \rangle\langle \Phi (0) | \right]
 \label{E:DE_def}
\end{eqnarray}
 is known as the AADM of the energy.  Thus, I conclude that
\begin{equation}
 \tau \geq \tau_\text{\tiny C} \equiv \frac{\left( 1 - \sqrt{\epsilon} \right)
 \hbar}{A \sum_j |\alpha_j|^2 | E_j - M |} \equiv \frac{\left( 1 -
 \sqrt{\epsilon} \right) \hbar}{A \,\,{\mathscr D}E} \equiv
 \frac{g_\text{\tiny C}(\epsilon) \hbar}{A \,\,{\mathscr D}E} .
 \label{E:time_bound_strengthened}
\end{equation}

\subsection{The Mixed State Case \label{Subsec:Mixed_State}}
 To extend the above bound to cover the case of mixed states, one simply needs
 to repeat the argument used by Giovannetti \emph{et al.} in
 Ref.~\cite{q_limit}: One can always purify the initial and final mixed states.
 And one may consider a particular choice of the purified states such that the
 two sets of orthonormal state kets of the ancillary systems used in the
 purification are identical.  Clearly, for other choice of the purified states,
 the evolution time $\tau$ can never be shorter than the above choice.  In
 addition, the fidelity between this pair of particularly chosen purified
 states does not exceed the fidelity between the original pair of mixed states.
 By applying the time bound in Eq.~(\ref{E:time_bound_strengthened}) to this
 particular choice of purified states, one concludes that the bound is also
 applicable to mixed states~\cite{q_limit}.

\begin{widetext}
\begin{table*}[t]
 \centering
 \begin{tabular}{||c|c|c|c|c||}
  \hline
  $|\Phi (0)\rangle$ & $\tau$ & $\tau_\text{\tiny TEUR}$ &
   $\tau_\text{\tiny ML}$ & $\tau_\text{\tiny C}$ \\
  \hline
  $\frac{1}{\sqrt{2}} \left( |-{\mathcal E} \rangle + |{\mathcal E}\rangle
   \right)$ & $\frac{\pi\hbar}{2{\mathcal E}}$ & $\frac{\pi\hbar}{2
   {\mathcal E}} = \tau$ & $\frac{\pi\hbar}{2{\mathcal E}} = \tau$ &
   $\frac{\hbar}{A {\mathcal E}} \approx 0.879\tau$
   \\
  \hline
  $|\varphi(0)\rangle$ ~as defined in~Eq.~(\ref{E:example_state}) &
   $\frac{\hbar}{A \alpha {\mathcal E}}$ & $\frac{\pi\hbar}{2\sqrt{\alpha}
   {\mathcal E}} \approx 0.876 \tau$ & $\frac{\pi \hbar}{2 {\mathcal E}}
   \approx 0.674 \tau$ & $\frac{\hbar}{A \alpha {\mathcal E}} = \tau$ \\
  \hline
  $|\varphi(0)\rangle$ ~as defined in~Eq.~(\ref{E:example_state}) but &
   $\frac{7\pi\hbar}{12 {\mathcal E}}$ & $\frac{\pi\hbar}{\sqrt{4-\sqrt{2}+
   \sqrt{6}} {\mathcal E}}$ & $\frac{\pi\hbar}{2 {\mathcal E}}$ &
   $\frac{4\hbar}{A {\mathcal E}}$
  \\
  with $\alpha = \frac{4}{4 - \sqrt{2} + \sqrt{6}} \approx 0.794$ & & $\approx
   0.764 \tau$ & $\approx 0.857 \tau$ & $\approx 0.598 \tau$
  \\
  \hline
  $\frac{1}{\sqrt{3}} \left( |0\rangle + |-{\mathcal E}\rangle +
   |{\mathcal E}\rangle \right)$ & $\frac{2\pi\hbar}{3{\mathcal E}}$ &
   $\frac{\pi\hbar}{{\mathcal E}} \sqrt{\frac{3}{8}} \approx 0.919 \tau$ &
   $\frac{\pi\hbar}{2 {\mathcal E}} = 0.75 \tau$ & $\frac{3\hbar}{2 A
   {\mathcal E}} \approx 0.988 \tau$ \\
  \hline
  $\frac{1}{\sqrt{2n}} \sum_{k=0}^{n-1} \left[ |-(k+\frac{1}{2})
   {\mathcal E}\rangle + |(k+\frac{1}{2}) {\mathcal E}\rangle \right]$ &
   $\frac{\pi\hbar}{n {\mathcal E}}$ & $\left( n \sqrt{\frac{3}{4n^2-1}}
   \right) \tau$ & $\left( \frac{n}{2n-1} \right) \tau$ & $\left( \frac{2}{A
   \pi} \right) \tau$ \\
  & & $\approx 0.866 \tau$ for large $n$ & $\approx 0.5\tau$ for large $n$ &
   $\approx  0.879 \tau$ for large $n$ \\
  \hline
  $\frac{1}{\sqrt{2n+1}} \sum_{k=-n}^n |k {\mathcal E} \rangle$ &
   $\frac{2\pi\hbar}{(2n+1) {\mathcal E}}$ & $\left[ \frac{2n+1}{4}
   \sqrt{\frac{3}{n(n+1)}} \right] \tau$ & $\left[ \frac{2n+1}{4n} \right]
   \tau$ & $\left[ \frac{(2n+1)^2}{2n(n+1)\pi A} \right] \tau$ \\
  & & $\approx 0.866\tau$ for large $n$ & $\approx 0.5\tau$ for large $n$ &
   $\approx 0.879 \tau$ for large $n$ \\
  \hline
 \end{tabular}
 \caption{Comparison between the three lower bounds on $\tau$ for $\epsilon =
  0$.
  \label{T:compare}
 }
\end{table*}
\end{widetext}

\subsection{Tightness Of The Bound \label{Subsec:Tightness}}
 The time bound in Eq.~(\ref{E:time_bound_strengthened}) is certainly tight
 when $\epsilon = 1$.  Hence, to show that this bound is tight for all
 $\epsilon$, I need only to consider the case of $\epsilon < 1$.  Let me
 consider the state
\begin{equation}
 |\varphi (0) \rangle = \sqrt{1-\alpha} |0\rangle + \sqrt{\frac{\alpha}{2}}
 |-{\mathcal E}\rangle + \sqrt{\frac{\alpha}{2}} |{\mathcal E}\rangle ,
 \label{E:example_state}
\end{equation}
 where
\begin{equation}
 \alpha = \frac{1-\sqrt{\epsilon}}{A x_m} \approx 0.592011 \left( 1 -
 \sqrt{\epsilon} \right) \in [0,1] .
 \label{E:alpha_def}
\end{equation}
 Furthermore, $|0\rangle$, $|{\mathcal E}\rangle$ and $|-{\mathcal E}\rangle$
 are normalized energy eigenkets with energies $0$, ${\mathcal E}$ and
 $-{\mathcal E}$, respectively.  Note that $\langle \varphi (0) | \varphi (t)
 \rangle = 1 - \alpha + \alpha \cos ( {\mathcal E} t / \hbar )$ is a
 real-valued sinusoidal function of $t$.  Besides, it starts to decrease at $t
 = 0$ until $t = \pi\hbar / {\mathcal E}$.  Therefore, the earliest time $\tau$
 at which $F(|\Phi (0)\rangle, |\Phi (\tau)\rangle) = | \langle \Phi (0) | \Phi
 (\tau) \rangle |^2 \leq \epsilon$ obeys
\begin{equation}
 \sqrt{\epsilon} = 1 - \alpha + \alpha \cos \left( \frac{{\mathcal E}
 \tau}{\hbar} \right) = 1 - \alpha + \alpha \cos \left( \frac{\tau
 \,\,{\mathscr D}E}{\alpha\hbar} \right) .
 \label{E:example_time_1}
\end{equation}
 From Eq.~(\ref{E:alpha_def}), I arrive at
\begin{eqnarray}
 \left( 1 - \sqrt{\epsilon} \right) \cos \left[ \frac{\tau A x_m
  \,\,{\mathscr D}E}{\left( 1 - \sqrt{\epsilon} \right) \hbar} \right] & = &
  \left( 1 - \sqrt{\epsilon} \right) (1 - A x_m) \nonumber \\
 & = & \left( 1 - \sqrt{\epsilon} \right) \cos x_m .
 \label{E:example_time_2}
\end{eqnarray}
 Note that I have used the fact that the line $y = A x$ intersects with the
 curve $y = 1 - \cos x$ at $x = x_m$ to arrive at the last line of the above
 equation.  Since $\epsilon < 1$, the general solution of
 Eq.~(\ref{E:example_time_2}) is

\begin{equation}
 \frac{\tau A \,\,{\mathscr D}E}{\left( 1 - \sqrt{\epsilon} \right) \hbar} = 1
 + \frac{2n\pi}{x_m}
 \label{E:example_time_3}
\end{equation}
 for all $n \in {\mathbb Z}$.  From Eq.~(\ref{E:x_m_numerical}), I know that
 $2\pi / x_m > 1$.  Therefore, the earliest time $\tau$ at which $| \langle
 \varphi (0) | \varphi (\tau) \rangle |^2 = \epsilon$ obeys $\tau A
 \,\,{\mathscr D}E / \left[ \left( 1 - \sqrt{\epsilon} \right) \hbar \right] =
 1$.  Thus, the bound stated in Eq.~(\ref{E:time_bound_strengthened}) is tight.
 After all the discussions above, it is clear that the maximum speed of
 evolution of a quantum state under the action of a time-independent
 Hamiltonian $H$ is tightly upper-bounded by $A \,\,{\mathscr D}E / [ (1 -
 \sqrt{\epsilon}) \hbar ]$.  And since the reciprocal of the speed of evolution
 of a quantum system signifies its quantum information processing
 rate~\cite{phy_limit,PHYSCOMP96,max_speed,q_limit}, the AADM of the energy
 ${\mathscr D}E$ is also a reasonable measure of the maximum possible quantum
 information processing rate of a system.

\section{Comparison With Existing Minimum Evolution Time Bounds
 \label{Sec:Compare}}
 Now, I start to compare the performance of the three bounds based on
 $\Delta E$, $E - E_0$ and ${\mathscr D}E$ for fixed values of $\epsilon$.
 Table~\ref{T:compare} shows the values of these three bounds when $\epsilon =
 0$ for a few cases in which $\tau$'s are known.  Clearly, the three bounds
 complement each other.  Moreover, $\tau_\text{\tiny C}$ is the best whenever
 ${\mathscr D}E / \Delta E$ and ${\mathscr D}E / (E - E_0)$ are small.  This
 finding is easy to understand.  From Eqs.~(\ref{E:TEUR_bound}),
 (\ref{E:ML_bound}) and~(\ref{E:time_bound_strengthened}), it is clear that for
 a fixed value of $\epsilon$, the performances of these three bounds is
 determined by the ratio $\Delta E : E-E_0 : {\mathscr D}E$.  And the
 $\tau_\text{\tiny C}$ bound works best when ${\mathscr D}E \ll \min ( \Delta
 E, E - E_0 )$.

 Observe that $E - E_0$ is the average absolute deviation from the ground state
 energy $E_0$.  (Consequently, the three bounds are in fact based on three
 different statistical dispersion measures of the eigenvalues of $H$ whose
 frequencies of occurrence are given by $|\alpha_j|^2$'s.)  So from our earlier
 discussions on AADM, ${\mathscr D}E \leq E - E_0$.  Furthermore, by a
 straight-forward application of the Cauchy-Schwarz inequality, one can show
 that ${\mathscr D}E \leq \Delta E$.  Note however that even though
 ${\mathscr D}E \leq \Delta E$ and $E - E_0$, it is still possible for the
 other two bounds to outperform Eq.~(\ref{E:time_bound_strengthened}) because
 the ratio $g_\text{\tiny TEUR}(\epsilon) : g_\text{\tiny ML}(\epsilon) : 2
 g_\text{\tiny C}(\epsilon) / \pi$ also plays a role in determining which bound
 is better.  But in any case, if the distribution formed by the eigenvalues of
 $H$ whose frequencies of occurrence are given by $|\alpha_j|^2$'s has a small
 kurtosis (whose value depends on $\epsilon$, of course), then the time bound
 due to ${\mathscr D}E$ is better than the other two.  As an illustration, I
 consider the special case in which $\epsilon = 0$ and the eigenvalues of $H$
 are drawn uniformly from an interval $[a,b]$.  The expected values of $\Delta
 E$, $E - E_0$ and ${\mathscr D}E$ are $(b-a)\sqrt{3}/6$, $(b-a)/2$ and
 $(b-a)/4$, respectively.  Thus, as the Hilbert space dimension of the state
 ket increases, the ratio $\tau_\text{\tiny TEUR} : \tau_\text{\tiny ML} :
 \tau_\text{\tiny C}$ approaches $\sqrt{3} : 1 : 4 / (\pi A) \approx 1.732 : 1
 : 1.757$ for a typical state ket $|\Phi (0)\rangle$.  So as a rule of thumb,
 $\tau_\text{\tiny C}$ has a good chance of giving a better time bound for
 $\tau$ when $\epsilon\approx 0$ provided that the kurtosis of the distribution
 of eigenvalues of $H$ is greater than or equal to the kurtosis of a uniform
 distribution, namely, $-6/5$.

 Finally, I study the effect of $\epsilon$ on the performance of the three
 bounds.  Note from Eqs.~(\ref{E:TEUR_bound}), (\ref{E:g2_def})
 and~(\ref{E:time_bound_strengthened}) that $g_\text{\tiny TEUR}(0) =
 g_\text{\tiny ML}(0) = g_\text{\tiny C}(0) = 1$.  Moreover, by differentiating
 $g_\text{\tiny C}(\epsilon) / g_\text{\tiny TEUR}(\epsilon)$ and
 $g_\text{\tiny C}(\epsilon) / g_\text{\tiny ML}(\epsilon)$ with respected to
 $\epsilon$, I conclude that $g_\text{\tiny C}(\epsilon) / g_\text{\tiny TEUR}
 (\epsilon)$ and $g_\text{\tiny C}(\epsilon) / g_\text{\tiny ML}(\epsilon)$ are
 decreasing and increasing functions of $\epsilon$, respectively.  In fact,
 $\lim_{\epsilon\rightarrow 0^+} g_\text{\tiny C}(\epsilon) /
 g_\text{\tiny TEUR}(\epsilon) = 0$.  In other words, for a sufficiently small
 value of $\epsilon$, it is likely that $\tau_\text{\tiny TEUR} \geq
 \tau_\text{\tiny C} \geq \tau_\text{\tiny ML}$.

\section{Conclusions \label{Sec:Final}}
 To summarize, I presented a tight lower bound $\tau_\text{\tiny C}$ for the
 time required to evolve between two states with fidelity less than or equal to
 $\epsilon$ under the action of a time-independent Hamiltonian.  And this
 time bound $\tau_\text{\tiny C}$ works best when the fidelity between the two
 states $\epsilon$ is small and the kurtosis of the distribution of eigenvalues
 of $H$ is $\gtrsim -6/5$.  My result also implies that the AADM of the energy
 ${\mathscr D}E$ is a reasonable measure of the maximum quantum information
 processing rate of a system.

\begin{acknowledgments}
 I thank C.-H.~F. Fung, K.~Y. Lee and H.-K. Lo for their discussions.  This
 work is supported by the RGC grant number HKU~700709P of the HKSAR Government.
\end{acknowledgments}

\bibliographystyle{apsrev}

\bibliography{qc46.2}

\end{document}